\documentclass[modern]{aastex7}
\newcommand\kms{\rm km~s$^{-1}$}

\begin{document}

\title{Detection of High-Velocity Na~I Absorption Toward the  Stellar Remnant of SN 1181 AD}


\author[orcid=0000-0003-4069-2817,gname='Peter', sname='Garnavich']{Peter Garnavich} 
\affiliation{Department of Physics and Astronomy, University of Notre Dame}
\email{pgarnavi@nd.edu}

\author[orcid=0000-0003-3829-2056,gname='Robert', sname='Fesen']{Robert A. Fesen} 
\affiliation{6127 Wilder Lab, Department of Physics and Astronomy, Dartmouth College, Hanover, NH 03755, USA}
\email{robert.fesen@dartmouth.edu }

\begin{abstract}

We report the detection of weak high-velocity \ion{Na}{1} 
absorption at V$_{\odot}$ = $-61.0\pm0.2$ km s$^{-1}$ in the spectrum of the stellar remnant at the center of the Galactic supernova remnant of 1181 AD. This velocity is not unlike that seen in old, more evolved SN remnants, but is much less than the remnant's $\simeq10^{3}$ km s$^{-1}$ expanding optical nebula. We briefly discuss its possible nature and origin.

\end{abstract}
 

\section{Introduction} 

WD~J005311 (J0053 hereafter) is believed to be the stellar remnant of the peculiar Galactic supernova of 1181 AD \citep{gvaramadze19,ritter21, schaefer23, lykou23}. The star is surrounded by an unusual nebula, Pa~30, visible at x-ray, optical, and infrared wavelengths but not in the radio \citep{oskinova20,ritter21,fesen23,shao25}. The supernova of 1181 is suspected to have been a Type~Iax event \citep{jha17} and its stellar remnant the merger of two white dwarf stars. Here we present the first high-resolution spectra of its stellar remnant at optical and near-infrared wavelengths.

\section{Observations} 

Spectra of J0053 were obtained with the Potsdam Echelle Polarimetric and Spectroscopic Instrument \citep[PEPSI;][]{pepsi15} on the Large Binocular Telescope (LBT). Observations were made on 2025 Nov. 14 and Dec. 19 UT. Exposure times were 2000~s using the CD4 cross-disperser that covers the wavelengths between 544~nm and 628~nm. All observations employed the D300 fiber that provides a resolving power of 50~000. In addition to the CD4 data, we obtained a single 2000~s exposure of J0053 with the CD6 cross-disperser covering $742-914$~nm. A short exposure of the bright A0 type star HIP~5518 was also obtained to mitigate the impact of telluric absorption features. The observations were processed by the PEPSI reduction pipeline resulting in continuum normalized spectra with wavelength corrected to heliocentric values. The optical spectrum around the Na~I D  lines (5890 \AA \ and 5896 \AA) and the K~I 7699 \AA \ line is shown in Figure~1.

\begin{figure*}[h!]
\plotone{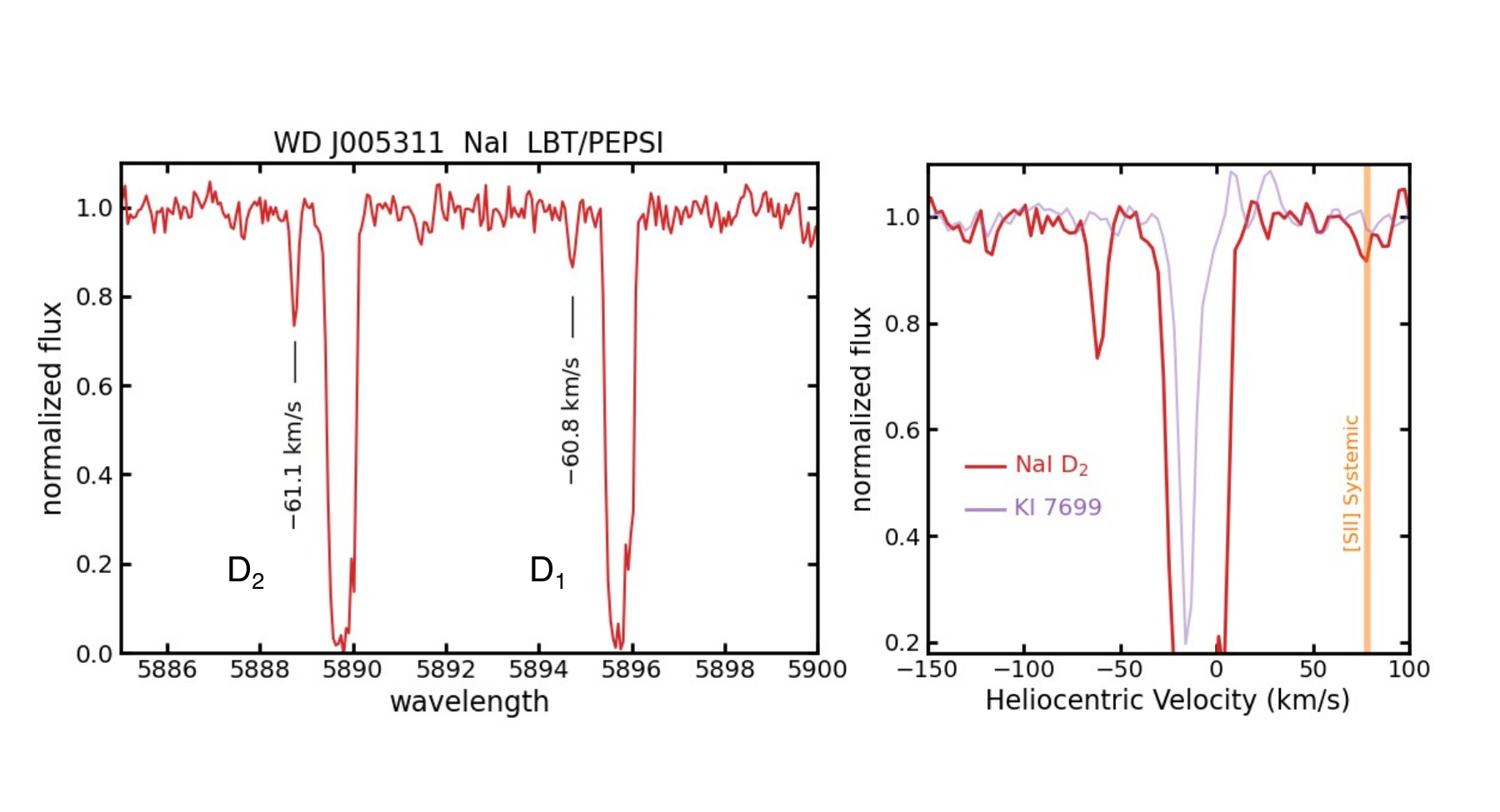}
\caption{{\bf Left:} High resolution spectrum of J0053, the star at the center of the SN 1181 remnant. Saturated interstellar absorption lines of Na~I are consistent with the significant dust extinction in the direction of J0053. High-velocity Na~I absorption is detected at $-61.0\pm0.2$~\kms\ in both Na~I components. {\bf Right:}  Close-up of the D$_2$ component (solid red line) plotted versus heliocentric velocity. The K~I absorption feature is shown as a light purple line. The high-velocity component is below the K~I detection limit. The solid orange band indicates the systemic velocity of the nebula's 
[\ion{S}{2}] emission assuming that it is expanding symmetrically.
\label{figure1}}
\end{figure*}

\section{Analysis} 

The spectrum of J0053 shows saturated Na~I D lines as expected given the A$_{0}$ = $2.8\pm 0.4$ mag extinction \citep{lykou23} toward the Galactic coordinates of $l=123.1^\circ$ and $b=4.6^\circ$. 
The centroids of the Na~I lines are uncertain given their saturation, but unsaturated K~I absorptions at 7665~\AA\ and 7699~\AA\ are seen in the near-infrared and are blueshifted by $-16\pm1$ \kms . Such a heliocentric velocity is as expected for this Galactic direction \citep{munch53}. The equivalent width (EQW) of the K~I 7699~\AA\ interstellar absorption is 230$\pm 10$~m\AA . This provides a reddening estimate of $E(B-V)=0.91\pm0.02$ based on the calibration by \citet{munari97} and this is consistent with the \citet{lykou23} estimate.

Relatively high-velocity Na~I absorptions is detected in both D components at a velocity of $-61.0\pm0.2$ km s$^{-1}$ but not in the near-infrared K~I lines. This is not surprising as the EQW of the 7699~\AA\ line is 10\%\ the D$_2$ line in the unsaturated regime \citep{munari97}.

The EQW of the high-velocity D$_2$ component is 45$\pm2$~m\AA\ and the D$_1$ component is 24$\pm2$~m\AA . The full-width at half-maximum (FWHM) of the high-velocity features is 162$\pm2$~m\AA , which is only slightly wider than the nominal instrument resolution.  
\citet{cunningham24} estimated the systemic velocity of the remnant to be $+78\pm4$ km s$^{-1}$ assuming the velocity of the nebula's [S~II] emission is spherically symmetric. If this velocity is adopted, then the detected Na~I is moving at $-139$~\kms\ relative to the centroid of the remnant.

\section{Discussion} 

High-velocity Na~I and Ca~II absorption lines with velocities of order $50 - 150$ km s$^{-1}$ have been detected toward several evolved Galactic supernova remnants (SNRs) using hot background stars or QSOs \citep[e.g.,][]{cha1999, fesen18, kochanek24, raymond24}. Ours is the first detection of high-velocity absorptions from a young SNR's stellar remnant outside that of the Crab Nebula's pulsar. However, since both the J0053 star and the Pa~30 remnant are unusual, the nature of this absorption and its location in the remnant are uncertain. Nonetheless, it seems unlikely that this $-60$~\kms\ gas lies within the SNR's optical nebula which, based on its [S~II] emission, has expansion velocities of 600 to 1400~\kms\  \citep{fesen23,cunningham24}. But it maybe that the absorption originates in the remnant between fallback and unbound ejecta \citep{ko24}.

Alternatively, the high-velocity  Na~I absorption could be the result of a mass loss wind driven shell from the WD merger if the subsequent supernova explosion did not occur within hours of the merger \citep{schwab12} but instead was delayed some $10^{4}$ yr  
until sufficient material has accreted to adiabatically compress the merged WD to high enough temperatures to ignite carbon, like that expected for lower mass mergers \citep{shen12}. In this case, the high-velocity Na~I gas we detected lies far outside the remnant's X-ray and optical emissions. With a possible velocity $\geq 100$ km s$^{-1}$, an outer ISM shock might be detectable through deep H$\alpha$ and 
[\ion{O}{3}] imaging.

\begin{acknowledgments}
We thank Ilya Ilyin for his work with the PEPSI data.

The LBT is an international collaboration among institutions in the United States, Italy and Germany. LBT Corporation Members are: The University of Arizona on behalf of the Arizona Board of Regents; Istituto Nazionale di Astrofisica, Italy; LBT Beteiligungsgesellschaft, Germany, representing the Max-Planck Society, The Leibniz Institute for Astrophysics Potsdam, and Heidelberg University; The Ohio State University, representing OSU, University of Notre Dame, University of Minnesota and University of Virginia. Observations have benefited from the use of ALTA Center (alta.arcetri.inaf.it) forecasts performed with the Astro-Meso-Nh model. Initialization data of the ALTA automatic forecast system come from the General Circulation Model (HRES) of the European Centre for Medium Range Weather Forecasts.
\end{acknowledgments}





%
\facilities{LBT}

\end{document}